\documentclass[
captions=tableabove
]{ceurart}

\usepackage{float}
\usepackage[parfill]{parskip}
\usepackage{caption}
\usepackage{subcaption}
\usepackage{lipsum}
\usepackage{graphicx}
\usepackage{placeins}

\usepackage{rotating}

\begin{document}

\copyrightyear{2022}
\copyrightclause{Copyright for this paper by its authors.
  Use permitted under Creative Commons License Attribution 4.0
  International (CC BY 4.0).}

\conference{AIXAS2022: Italian Workshop on Artificial Intelligence for an Ageing Society, Udine, Italy, November 29, 2022}

\title{Sentiment recognition of Italian elderly through domain adaptation on cross-corpus speech dataset}

\author[1]{Francesca Gasparini}[%
orcid=0000-0002-6279-6660,
email=francesca.gasparini@unimib.it
]
\author[1]{Alessandra Grossi}[%
orcid=0000-0003-1308-8497,
email=alessandra.grossi@unimib.it
]

\address[1]{Department of Computer Science, Systems and Communications, University of Milano - Bicocca, Italy}

\begin{abstract}
 The aim of this work is to define a speech emotion recognition (SER) model able to recognize positive, neutral and negative emotions in natural conversations of Italian elderly people. Several datasets for SER are available in the literature. However most of them are in English or Chinese, have been recorded while actors and actresses pronounce short phrases and thus are not related to natural conversation. Moreover only few speeches among all the databases are related to elderly people. Therefore, in this work, a multi-language and multi-age corpus is considered merging a dataset in English, that includes also elderly people, with a dataset in Italian. A general model, trained on young and adult English actors and actresses is proposed, based on XGBoost. Then two strategies of domain adaptation are proposed to adapt the model either to elderly people and to Italian speakers. The results suggest that this approach increases the classification performance, underlining also that new datasets should be collected.

\end{abstract}


\maketitle

\section{Introduction}
\label{sec:intro}
Emotions play a relevant role in defining individuals' behaviours and coordination in human-human interactions \cite{lange2022reading}. In particular, humans find speech conversations more natural and effective than its written form as way to express themselves \cite{el2011survey}. During conversations, people try to convey their thought not only by words but also by bodily, vocal or facial expressions \cite{lange2022reading, swain2018databases}. Specifically in vocal expressions the affective state of individuals is expressed both by the linguistic and acoustic information carried by the speech \cite{anagnostopoulos2015features}. For instance, the same sentence said with different intonations can express different emotions by the speaker and, thus, can lead to a different response from the listener \cite{wu2022design}. 
Therefore, in order to create a natural interaction between humans and computers, the machine must be able to understand emotions from the speaker's voice and consequently adapt. Speech Emotion Recognition (SER) consists of the task of processing and classifying speech signals in order to recognize the emotional state of the speaker \cite{wu2022design, akccay2020speech}. Systems based on SER have different fields of application, such as health care \cite{france2000acoustical}, e-learning tutoring \cite{hua2006using}, automotive \cite{cevher2019towards} or entertainment \cite{nakatsu1999emotion, alhargan2017multimodal}. In particular, these kinds of systems can be employed for the definition of diagnostic tools able to help therapists in detecting psychological disorders \cite{low2010detection} or for automatically recognising mental state alteration in drivers \cite{al2011novel}. Automatic emotion detection systems can also be used in the call center or mobile communications to detect the emotions of callers and to help agents improving the quality of service \cite{gupta2007two, vaudable2012negative}, or in human-robot interactions to support a more natural and social communication between human and machine \cite{hegel2006playing, jones2008affective}.\newline
Several researches have been carried out in the field of Speech Emotion Recognition during the last three decades \cite{fahad2021survey}. In particular, many of these analysis are performed considering only one between linguistic or acoustic information of speech while in recent analysis a multi-modal approach is examined \cite{atmaja2022survey}.

In our study, we focus only on acoustic information. In this field, both traditional machine learning and deep learning approaches have been taken into account in previous literature. In general, the traditional pipeline in a SER system consists of three steps: signal preprocessing, features extraction and classification \cite{thakur2021speech}. Concerning features extractions, different set of features have been tested: traditional features extracted by audio signals \cite{el2011survey}, including prosodic (such as pitch, energy and duration), spectral (such as fundamental frequency, Mel Frequency Cepstral Coefficients or Linear Prediction Cepstral Coefficients) and voice quality features (such as jitter or shimmer), as well as deep features extracted by pre-trained networks. In this latter, the audio signals are usually represented as Spectrogram or Scalogram and used as input to pre-trained network to extract features \cite{stolar2017real, boateng2020speech}. 
With reference to classifiers, in several research such as \cite{swain2015study, latif2018cross}, traditional classifiers have been employed. In particular, according to \cite{wani2021comprehensive}, the classical classification techniques preferred in SER system are Gaussian Mixture Model, Hidden Markov Model, Artificial Neutral Network, Decision Trees and Support Vector Machine. In few analysis \cite{lugger2009combining, schuller2005robust} also ensemble techniques combining several classifiers have been tested. Deep approaches have been also considered in the last years. In particular, framework using Convolutional Neutral Network (CNN) \cite{badshah2017speech}, Recurrent Neural Network (RNN) \cite{aghajani2020speech} and Long Short-Term memory network (LSTM) \cite{cho2019deep} have been evaluated, using both traditional features \cite{kumbhar2019speech} and raw audio signals. In some cases, also mechanism of attention \cite{atmaja2019speech, jian2022speech} or auto encoding \cite{neumann2019improving} have been added to classifiers in order to increase performance. The main SER approaches have been summarized in review manuscripts such as \cite{akccay2020speech} or \cite{wani2021comprehensive}. \newline 
Despite the huge number of analyzes carried out, there are still numerous issues that make difficult to recognize emotions in speech. In \cite{fahad2021survey}  some of these challenges and the approaches tested so far to solve them are summarized. In particular, speech emotion recognition algorithms struggle in recognize emotions when people of different language or age are considered. \newline
In literature there are many datasets collected for SER purpose. These corpora can be classified into three groups with reference to how emotional speech is generated \cite{koolagudi2012emotion}: i) Acted datasets, where the data are collected from actors/actresses that try to simulate emotions; ii) Evoked or Elicited datasets, where the subjects are involved into situations especially created to evoke or induce certain emotions; and iii) Spontaneous or Natural datasets, which contain more authentic emotions as collected from real-world situations like call-centers or public places \cite{fahad2021survey}. 
Most of the datasets available in the literature are composed of recited speeches \cite{ringeval2013introducing}, while only few of them consider natural conversations \cite{steidl2009automatic, fan2021lssed, morrison2007ensemble}. 
Moreover, the considered languages are mainly English and Chinese. It has been demonstrated that language has a strong influence in how emotions are expressed \cite{latif2018cross}, and thus multi-language datasets have been proposed \cite{parada2018categorical, hozjan2002interface}.
Age is another factor that influences the acoustic characteristics of the voice, especially in the case of elderly \cite{deliyski2001effects, sundberg1998age}. However, this is still an open field of research and few works face the problem of SER in case of elderly, or varying the age \cite{boateng2020speech, jian2022speech, verma2016age, souganciouglu2020everything}, and old subjects are rarely present in available datasets \cite{schuller2020interspeech, cao2014crema, SP2/E8H2MF_2020, fan2021lssed}.  

In this work we consider the problem of SER, considering elderly Italian people. Moreover we focus on positive, neutral and negative emotions.
We propose to consider a multi-language, multi-aged approach, considering a cross-corpus dataset, described in Section \ref{chap:dataset}. We start from a general model trained on an English dataset of young and adult subjects, and we refine this model to adapt either to elderly and Italian language, as described in Section \ref{chap:DA strategies}, adopting two different domain adaptation techniques. In Section \ref{chap:preprocess and fe} preprocessing of raw data, feature extraction and data augmentation, needed to apply the proposed solutions are presented.
The results, discussed in Section \ref{chap:results}, underline the potentialities and the limits of the proposed approaches, while future perspective are drawn in the Conclusions.

\section{Cross-corpus dataset}
\label{chap:dataset}
 
In this work, we consider two  datasets available in the literature, labeled with emotions, and characterized by the presence of elderly subjects or by the presence of Italian sentences: the CRowd-sourced Emotional Multimodal Actors dataset CREMA-D \cite{cao2014crema} and EMOVO \cite{costantini2014emovo}.  \newline\newline
\textit{CREMA-D} \cite{cao2014crema} is a free audio-visual dataset collected to investigate facial and vocal expressions and perception of acted emotions. It consists of 7442 audio and video recordings of professional actors playing 12 utterances each one expressed in six emotional states (happy, sad, anger, fear, disgust and neutral) at different intensity levels. In the first utterance, the actors were directed to simulate each emotion in three levels of intensity (low, medium and high) while, for the other eleven sentences, they were free to express the emotion at their preferred intensity. The sentences selected for the experiment are in English and have a neutral semantic content. In total, 48 actors and 43 actresses of different ages and ethnicity were involved in the experiments, including 6 elderly with more than 60 years and 85 adults aged between 20 and 59 years. For the purpose of our analysis, the two groups of subjects are considered separately with a total of 492 signals for elderly, named hereinafter \textit{CREMA-D-ELD}, and 6950 signals for adults (\textit{CREMA-D-ADULT}). For further details of CREMA-D dataset, please refer to the reference manuscript \cite{cao2014crema}. \newline\newline
\textit{EMOVO} \cite{costantini2014emovo} is an acted free audio speech emotional dataset based on the Italian language. The corpus was collected from six young Italian actors (3 male and 3 female) with a mean age of 27.1 (no elderly actors were involved). Similarly to CREMA-D, in the experimental protocol, 14 utterances had to be performed by the actors simulating different emotional states. In particular, for each utterance, 7 affective states were considered: neutral, disgust, fear, anger, joy, surprise and sadness. The total number of utterances collected in the dataset is 588, with a mean of 98 signals per actor. More details about EMOVO can be found in \cite{costantini2014emovo}\newline\newline
In Table \ref{tab:dataset_info}  the main information about these two datasets are summarized.

\begin{table}[!tbp]
  \centering
  \caption{Summary of main CREMA-D and EMOVO characteristics}
   \hspace*{-1cm}
  \begin{tabular}{|p{4.5em}p{2em}p{7em}cp{4em}p{5em}p{3em}p{3em}p{3em}|}
    \toprule
    \toprule
    \multicolumn{1}{|p{4em}}{\centering \textbf{Dataset Name}} & \centering
    \textbf{Type} & \multicolumn{1}{p{6em}}{ \textbf{Emotions Considered}} & \textbf{Language} &  \textbf{No. of \newline Utterances} & \centering \textbf{Tot. No. of \newline Subjects} & \centering \textbf{No. of \newline Males} & \centering \textbf{No. of \newline Elderly} & \textbf{Mode} \\
    \midrule
    CREMA-D & Acted & happy, sad, anger, fear, disgust and neutral & \multicolumn{1}{c}{\small English} & \multicolumn{1}{c}{\normalsize 12} & \multicolumn{1}{c}{\normalsize 91} & \multicolumn{1}{c}{\normalsize 48} & \multicolumn{1}{c}{\normalsize 6}  & \multicolumn{1}{p{3em}|}{\small Audio/ \newline visual} \\
     &  &  &  &  &  & &  & \\
    EMOVO & Acted & joy, surprise, sad, anger, fear, disgust and neutral & \multicolumn{1}{c}{\small Italian} & \multicolumn{1}{c}{\normalsize 14} & \multicolumn{1}{c}{\normalsize 6}  & \multicolumn{1}{c}{\normalsize 3}  & \multicolumn{1}{c}{\normalsize 0}  & \small Audio \\
    \bottomrule
    \bottomrule
    \end{tabular}%
  \label{tab:dataset_info}%
\end{table}%

\begin{figure}[!tb]
    \begin{minipage}{0.50\textwidth}
        \centering
        \includegraphics[width=1\linewidth]{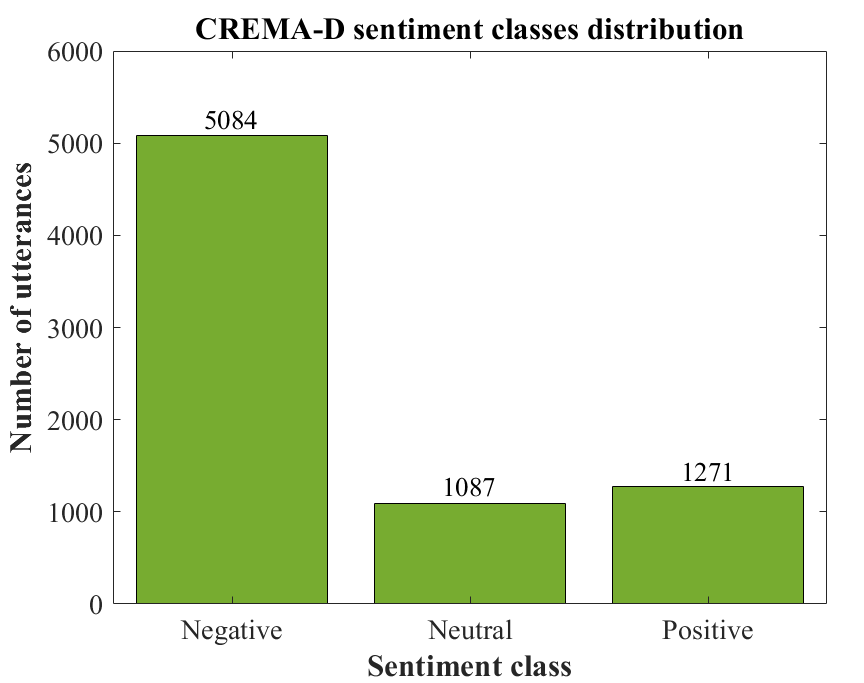}
    \end{minipage}\hfill
    \begin{minipage}{0.50\textwidth}
        \centering
        \includegraphics[width=1\linewidth]{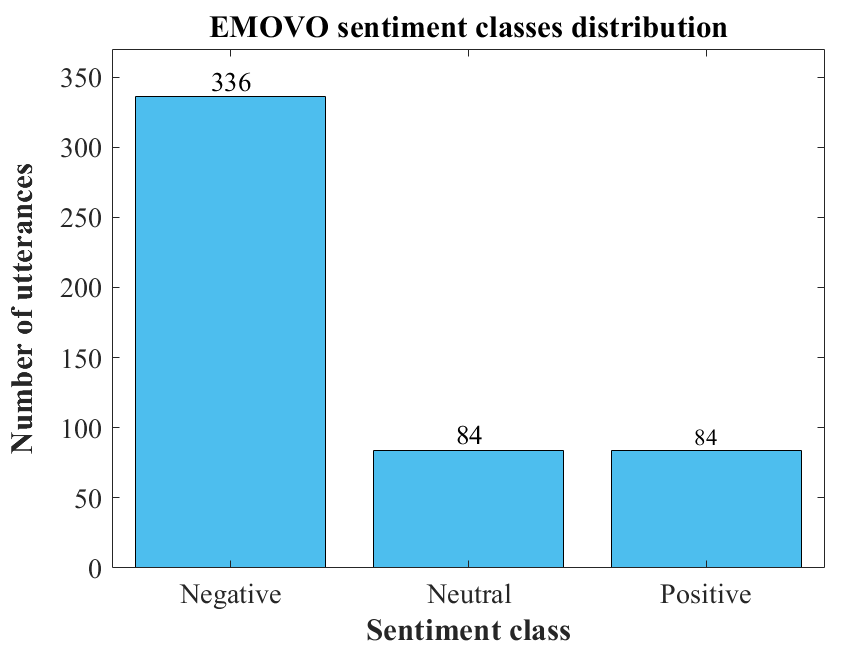}
    \end{minipage}
    \caption{Number of utterances mapped as negative, positive or neutral in the two datasets CREMA-D (left) and EMOVO (right)}
    \label{fig:class_distribution}
\end{figure}

In both the selected datasets, the signals are labeled using the six basic emotions defined by Ekman. In order to use these datasets in our analysis, each emotion has been converted into its respective sentiment according with the mapping defined in \cite{poria2018meld}. In particular, we have considered anger, fear, disgust and sadness as negative sentiments, happy (or joy) as positive sentiment and neutral as neutral sentiment. All the EMOVO signals labeled as ``surprise'' has been instead excluded from the analysis as difficult to be mapped into a single sentiment class \cite{poria2018meld}. The distribution of the utterances in the three sentiment classes is shown in Figure \ref{fig:class_distribution} for the two datasets considered. 

Concerning the sentiment analysis, two other datasets are usually adopted in Speech Sentiment Recognition researches: Multimodal EmotionLines Dataset (MELD) \cite{poria2018meld} and CMU Multimodal Opinion Sentiment and Emotion Intensity (CMU-MOSEI) \cite{zadeh2018multimodal}. The first \cite{poria2018meld} is a data corpus composed by more then 13000 utterances from 1433 dialogues from the TV-series Friends and labeled with three sentiment class: negative, positive and neutral. CMU-MOSEI \cite{zadeh2018multimodal}, instead, contains 23453 annotated video-clips from 250 different topics, gathered from online video sharing websites and labeled with sentiment in Likert scale. Despite both the datasets are directly labeled with sentiment, they were excluded from our analysis. In particular, concerning MELD, the dataset has been discarded due to the presence, in several audios, of laugh tracks or multiple voices overlapping the main actor's speech. This makes the audio signal very noisy and makes it difficult to identify which part of the audio is related to the labelled sentiment. With reference to CMU-MOSEI, instead, the dataset has been excluded from the study because of the lack of the subject's age that makes impossible to separate signals collected from elderly from the one's collected from young or adults.\newline

\section{Data Adaptation strategies}
\label{chap:DA strategies}
The proposed analysis considers two research hypotheses:
\begin{itemize}
    \item \textit{Domain adaptation based on age}, training a general Speech Sentiment Recognition model using speech data collected from English young and adults subjects and adapting this model on new data collected from English elderly subjects. 
    \item \textit{Domain adaptation based on language}, trying to refine a pre-trained Speech Sentiment Recognition model on English young and adults subjects to recognize new data collected from Italian young and adults people. 
\end{itemize}
In all the experiments performed, the gradient boosted decision trees algorithm implemented as XGBoost \cite{chen2016xgboost} has been selected as classification model while two different instance weighting domain adaptation strategies have been tested: 
\begin{itemize}
    \item the \textit{Kullback-Leiber Important Estimation Procedure (KLIEP) strategy} \cite{sugiyama2007direct} that assigns a weight to the training instances during the classifier learning task in order to minimize the Kullback-Leibler divergence between train and target distributions. In our analysis we have considered the supervised implementation of this algorithm using  ``rbf'' as Kernel with two different gamma: 0.1 and 1.  
    \item \textit{the Transfer AdaBoost for Classification (TrAdaBoost)} \cite{inproceedings} is a supervised domain adaptation strategy that extends boosting-based learning algorithms to the field of transfer learning. In particular, at each iteration, the algorithm trains a new weak classifier giving less importance to the training instances poorly predicted in previous iterations while emphasising the target samples correctly recognized. The final model is the combination of the last half computed estimators weighted according to their relevance. The number of iterations selected in our experiments is 10.
\end{itemize}
The application of these two strategies requires to split the data into three distinct sets: i) Training (or Source) set made up of a large amount of labeled data used to train the general model; ii) Target set consisting of few samples belonging to a new but related domain that are used to adapt the general model to this new data distribution and iii) Test set composed by data similar to Target set and used to evaluate the model performances. In our experiments, the definition of these three sets changes according to the research hypothesis considered. In particular, in multi-age analysis, the data of CREMA-D-ADULT have been used as Training set while Target and Test sets have been defined as subsets of CREMA-D-ELD. Instead, in multi-language analysis, the training of the general model is performed using CREMA-D-ADULT data while Target and Test sets are both defined as partitions of EMOVO data.

Different validation strategies have been tested to partition the data of CREMA-D-ELD and EMOVO into Target and Test set:
\begin{itemize}
    \item Leave One Subject Out (LOSO) Cross Validation strategy, where the folds are partitioned according to subject and thus, at each iteration, all the data of a single subject are used as Test set while the data of the remaining subjects are used as Target set.
    \item Leave One Utterance Out (LOUO) Cross Validation strategy, where the folds are defined according to the pronounced utterances, thus at each iteration, all the data related to a single utterance are used to test the model while the data of the remaining utterances are used as Target set.
\end{itemize}
To test the performances of our classification models, several well-known evaluation metrics are computed \cite{grandini2020metrics} including the accuracy, single class F1-score, evaluated as the harmonic mean of single class precision and recall, and macro F1-score \cite{lipton2014optimal} computed as the unweighted mean of the single class F1-score.

\section{Model input data}
\label{chap:preprocess and fe}

\begin{figure}[!tb]
    \centering
    \includegraphics[width=1\linewidth]{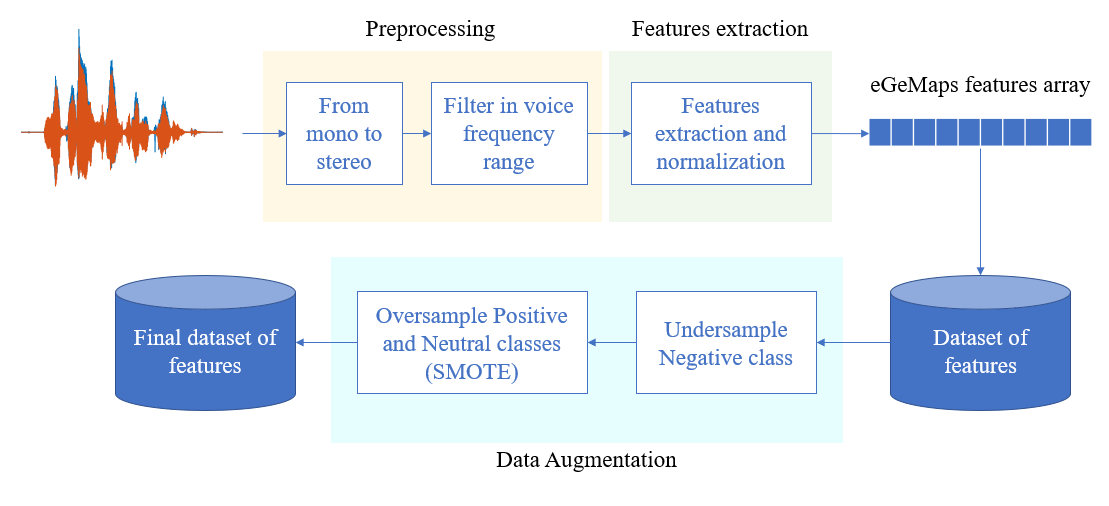}
    \caption{Pipeline used, in the analysis, for extracting features from the signals of Training and Target datasets. Concerning the Test set, the Data Augmentation step is not applied.}
    \label{fig:pipeline_audio}
\end{figure}

To apply the strategies of domain adaptation described in the previous section, preprocessing, feature extraction, and data augmentation to balance the classes have been performed on raw data. The whole process is depicted in Figure \ref{fig:pipeline_audio}.

\subsection{Preprocessing}
The audio signals of each dataset are preprocessed to extract only the information concerning the target speaker's voice. In particular, the audio clips were first converted from stereo to mono by averaging samples across the two channels. Then, each signal was filtered  using a pass-band Butterworth filter with lower cutoff frequency at 300 Hz and upper cutoff frequency at 3000 Hz to removes the spectral components out of the voice frequency range \cite{birch2021environmental}. \newline

\subsection{Feature extraction}
From the pre-processed signals, the eGeMAPS acoustic feature set was extracted using the python library implementation of openSMILE toolkit \cite{eyben2010opensmile}. The eGeMAPS feature set (extended Geneva Minimalistic Acoustic Parameter Set) \cite{eyben2015geneva} is a set of audio features proposed for affective analysis in voice signals. 
It consists of 25 Low Level Descriptor (LLD) features including energy, frequency, cepstral, spectral and dynamic parameters. In order to summarize the variation of these parameters over the time windows, some high level functional features are extracted using statistical functions as arithmetic mean, standard deviation or percentile. Applying these statistics, a total of 88 features have been extracted for each considered signal. 
The extracted features have been normalized by z-scoring in order to reduce inter signals differences. \newline

\subsection{Data Augmentation}

Only for Training (or Source) and Target dataset, the feature extraction step has been followed by data augmentation.
In both the datasets, the cardinality of the negative sentiment class is four times greater then positive or neutral ones. This is due to an imbalance among the number of emotions mapped as negative (angry, fear, sadness, disgust) and the number of emotions mapped as positive (happy) and neutral (neutral) in the selected emotion-sentiment transformation. In order to create more balanced classes, a two steps procedure have been applied to training and target data according to the experiment considered . First the majority class have been under-sampled, discarding randomly half of the negative instances. In this process, the discarded elements have been selected trying to keep balanced the number of elements for each negative emotions. Then an oversampling strategy based on SMOTE  algorithm \cite{chawla2002smote} has been applied to increase the number of samples in the two minority classes (positive and neutral). SMOTE (Synthetic Minority Oversampling TEchnique) \cite{chawla2002smote} is an oversampling method that random generates new synthetic data for the minority class starting from the original data points. In particular, at each iteration, the algorithm selects one of the k-nearest neighbors of a random minority class element and create new artificial elements linear interpolating the two instances using a random number between zero and one. The procedure is repeated until the cardinality of the classes is balanced.

\section{Results and discussion}
\label{chap:results}
The aim of this work is define a classification model able to automatically recognize three sentiment states (positive, neutral and negative) using acoustic features extracted from speech when different age and language are considered. In particular, two different experiments have been carried out to evaluate the research hypothesis described in Section \ref{chap:DA strategies}: domain adaptation on elderly and domain adaptation on language.

\subsection{Domain adaptation on elderly}

\begin{table}[tbp]
  \centering
  \caption{Experiments carried out in cross-age analysis varying the Domain Adaptation Strategy considered (second column) and the Validation strategy adopted (last column). In all the experiments, the general model has been trained using the data of CREMA-D-ADULT while different subsets of CREMA-D-ELD dataset have been selected as Target and Test set.}
  \hspace*{-1.2cm}
  \renewcommand{\arraystretch}{1.4}
    \begin{tabular}{|c|p{6em}|p{7em}|l|l|p{6em}|}
    \toprule
    \toprule
    & \multirow{2}{*}{\textbf{\normalsize DA Strategy}} & \multicolumn{1}{c|}{\multirow{2}{*}{\textbf{\normalsize Training Set}}} & \multicolumn{1}{c|}{\multirow{2}{*}{\textbf{\normalsize Target set}}} & \multicolumn{1}{c|}{\multirow{2}{*}{\textbf{\normalsize Test set}}} & \multicolumn{1}{p{6em}|}{\centering \textbf{\normalsize Validation Strategy}} \\
    \cline{1-6}
    \multirow{6}[6]{*}{\begin{sideways}XGBoost classifier\end{sideways}} &    & \multicolumn{1}{c|}{\textit{CREMA-D-ADULT}} & \multicolumn{2}{c|}{\textit{CREMA-D-ELD}} &  \\
    \cline{2-6}    
     & No Domain Adaptation & 
    \multicolumn{1}{p{7em}|}{all 85 young and adults subjects} & \multirow{2}{*}{no target dataset} & \multirow{2}{*}{6 elderly x 12 utterances} & \multicolumn{1}{p{6em}|}{\centering Training / Test independent} \\
    \cline{2-6} &     
    \multirow{2}[1]{*}{KLIEP} & 
    \multirow{2}[1]{7em}{all 85 young and adults subjects} & 5 elderly x 12 utterances  & 1 elderly x 12 utterances  & \multicolumn{1}{p{6em}|}{\centering LOSO} \\ [-1.7em] &    &    &  &  &  \\
    &    &    & 11 utterances x 6 elderly & 1 utterance x 6 elderly & \multicolumn{1}{p{6em}|}{\centering LOUO} \\
    \cline{2-6}       
    & \multirow{2}[2]{*}{TrAdaBoost} & \multirow{2}[2]{8em}{all 85 young and adults subjects} & 5 elderly x 12 utterances  & 1 elderly x 12 utterances  & \multicolumn{1}{p{6em}|}{\centering LOSO} 
    \\ [-1.7em] &    &    &  &  &  \\
     &    &    & 11 utterances x 6 elderly & 1 utterance x 6 elderly & \multicolumn{1}{p{6em}|}{\centering LOUO} \\
    \bottomrule
    \bottomrule
    \end{tabular}%
  \renewcommand{\arraystretch}{1}
  \label{tab:expe_CREMA_D}%
\end{table}%

\begin{table}[tbp]
  \centering
  \caption{Cross-age performance comparison using CREMA-D-ADULT as training set and CREMA-D-ELD as target and test set. The analysis are performed varying the Domain Adaptation strategy (second column) and Performance Evaluation method (third column). Three evaluation metrics are considered: macro F1-score, accuracy and single class F1-score.}
  \hspace*{-0.8cm}
  \renewcommand{\arraystretch}{1.4}
    \begin{tabular}{|c|cc|c|ccc|c|}
    \toprule
    \toprule
    \textbf{Classifier} & \textbf{DA Strategy} & \multicolumn{1}{p{6em}|}{\centering \textbf{Validation Strategy}} & \multicolumn{1}{p{4em}|}{\centering \textbf{Macro F1-score}} & \multicolumn{1}{p{4.22em}}{\textbf{Negative F1-score}} & \multicolumn{1}{p{3.945em}}{\textbf{Neutral F1-score}} & \multicolumn{1}{p{3.945em}|}{\textbf{Positive F1-score}} & \textbf{Accuracy} \\
    \hline
    \multirow{5}[6]{*}{\begin{sideways}XGBoost classifier\end{sideways}} & \multicolumn{1}{p{5.5em}}{\centering No Domain Adaptation} & \multicolumn{1}{p{6em}|}{\centering Training / Test independent} & \multirow{2}{*}{\normalsize 62\%} & \multirow{2}{*}{\normalsize 0,79} & \multirow{2}{*}{\normalsize 0,55} & \multirow{2}{*}{0,51} & \multirow{2}{*}{\normalsize 70\%} \\
    \cline{2-8}       
    & \multirow{2}[2]{*}{KLIEP} & LOSO & \normalsize 60\% & \normalsize 0,76 & \normalsize 0,57 & \normalsize 0,49 & \normalsize 67\% \\
    &    & LOUO & \normalsize 60\% & \normalsize 0,76 & \normalsize 0,56 & \normalsize 0,47 & \normalsize 67\% \\
    \cline{2-8}      
    & \multirow{2}[2]{*}{TrAdaBoost} & LOSO & \normalsize 62\% & \normalsize 0,79 & \normalsize 0,56 & \normalsize 0,52 & \normalsize 70\% \\
    &    & LOUO & \normalsize 62\% & \normalsize 0,78 & \normalsize 0,55 & \normalsize 0,52 & \normalsize 69\% \\
    \bottomrule
    \bottomrule
    \end{tabular}%
  \renewcommand{\arraystretch}{1}
  \label{tab:performance_DA_eld}%
\end{table}%

In the first analysis, a multi-age corpus sentiment classification is considered. As described in Section \ref{chap:DA strategies}, the two parts of CREMA-D dataset have been used respectively for Training set (CREMA-D-ADULT) and Target and Test set (CREMA-D-ELD). For each domain adaptation strategy, two different evaluation methods are tested: LOSO and LOUO. The results achieved in these experiments are compared with the performances reached by the XGBoost model when no domain adaptation strategy is applied. In this case, thus, the classifier is trained on CREMA-D-ADULT data and tested on the independent dataset CREMA-D-ELD. The classification settings considered in the analysis are summarized in Table \ref{tab:expe_CREMA_D}. For each of these analyses, Table \ref{tab:performance_DA_eld} reported the classification performance achieved by the XGBoost classifiers in terms of accuracy, macro F1-score and single class F1-score. The results show how, in case of elderly, the use of domain adaptation techniques does not significantly increase the performances of the classification model with reference to the benchmark case without adaptation. A macro F1-score value of 62\%, in fact, is achieved both when TrAdaBoost or no domain adaptation is applied. Lower performances are instead obtained using the KLIEP domain adaptation algorithm with F1-score value near to 60\%. Similar results are reached using both LOSO and LOUO evaluation strategies. \newline
Considering the values of per-class F1-scores reached emerges how, in all the experiments performed, the Negative class appears easier to be recognized than Neural and Positive ones. This difference can be due to the presence of a higher number of different instances in the negative class than in the other two classes where several instances were artificially created using SMOTE data augmentation strategy.
\newline
From these preliminary results, it seems that data adaptation does not increase the performance of the proposed SER model. This is probably related to several aspects. The elderly here considered are actors or actresses, and thus they are not so significantly different from a population of young and adult persons. Moreover the elderly are only 6, of which only one is a female. A more realistic dataset should be consider to proper verify this research question.

\subsection{Domain adaptation on language}

\begin{table}[tbp]
  \centering
  \caption{Experiments carried out in cross-language analysis varying the Domain Adaptation Strategy considered (second column) and the Validation strategy adopted (last column). In all the experiments the general model has been trained using the data of CREMA-D-ADULT while different subsets of EMOVO dataset have been selected as Target and Test set.}
    \hspace*{-1.2cm}
  \renewcommand{\arraystretch}{1.4}
    \begin{tabular}{|c|p{6em}|p{7em}|l|l|p{6em}|}
    \toprule
    \toprule
    & \multirow{2}{*}{\textbf{\normalsize DA Strategy}} & \multicolumn{1}{c|}{\multirow{2}{*}{\textbf{\normalsize Training Set}}} & \multicolumn{1}{c|}{\multirow{2}{*}{\textbf{\normalsize Target set}}} & \multicolumn{1}{c|}{\multirow{2}{*}{\textbf{\normalsize Test set}}} & \multicolumn{1}{p{6em}|}{\centering \textbf{\normalsize Validation Strategy}} \\
    \cline{1-6}
    \multirow{6}[6]{*}{\begin{sideways}XGBoost classifier\end{sideways}} &    & \multicolumn{1}{c|}{\textit{CREMA-D-ADULT}} & \multicolumn{2}{c|}{\textit{EMOVO}} &  \\
    \cline{2-6}    
     & No Domain Adaptation & 
    \multicolumn{1}{p{7em}|}{all 85 young and adults subjects} & \multirow{2}{*}{no target dataset} & \multirow{2}{*}{6 subjects x 14 utterances} & \multicolumn{1}{p{6em}|}{\centering Training / Test independent} \\
    \cline{2-6} &     
    \multirow{2}[1]{*}{KLIEP} & 
    \multirow{2}[1]{7em}{all 85 young and adults subjects} & 6 subjects x 14 utterances  & 1 subject x 14 utterances  & \multicolumn{1}{p{6em}|}{\centering LOSO} \\ [-1.7em] &    &    &  &  &  \\
    &    &    & 13 utterances x 6 subjects & 1 utterance x 6 subjects & \multicolumn{1}{p{6em}|}{\centering LOUO} \\
    \cline{2-6}       
    & \multirow{2}[2]{*}{TrAdaBoost} & \multirow{2}[2]{8em}{all 85 young and adults subjects} & 5 subjects x 14 utterances  & 1 subject x 14 utterances  & \multicolumn{1}{p{6em}|}{\centering LOSO} 
    \\ [-1.7em] &    &    &  &  &  \\
     &    &    & 13 utterances x 6 subjects & 1 utterance x 6 subjects & \multicolumn{1}{p{6em}|}{\centering LOUO} \\
    \bottomrule
    \bottomrule
    \end{tabular}%
  \renewcommand{\arraystretch}{1}
  \label{tab:expe_EMOVO}%
\end{table}%

\begin{table}[tbp]
  \centering
  \caption{Cross-language performance comparison using CREMA-D-ADULT as training set and EMOVO as target and test set. The analysis are performed varying the Domain Adaptation strategy (second column) and Performance Evaluation method (third column). Three evaluation metrics are considered: Macro F1-score, Accuracy and single class F1-score.}
  \hspace*{-0.8cm}
  \renewcommand{\arraystretch}{1.4}
    \begin{tabular}{|c|cc|c|ccc|c|}
    \toprule
    \toprule
    \textbf{Classifier} & \textbf{DA Strategy} & \multicolumn{1}{p{6em}|}{\centering \textbf{Validation Strategy}} & \multicolumn{1}{p{4em}|}{\centering \textbf{Macro F1-score}} & \multicolumn{1}{p{4.22em}}{\textbf{Negative F1-score}} & \multicolumn{1}{p{3.945em}}{\textbf{Neutral F1-score}} & \multicolumn{1}{p{3.945em}|}{\textbf{Positive F1-score}} & \textbf{Accuracy} \\
    \hline
    \multirow{5}[6]{*}{\begin{sideways}XGBoost classifier\end{sideways}} & \multicolumn{1}{p{5.5em}}{\centering No Domain Adaptation} & \multicolumn{1}{p{6em}|}{\centering Training / Test independent} & \multirow{2}{*}{\normalsize 35\%} & \multirow{2}{*}{\normalsize 0,71} & \multirow{2}{*}{\normalsize 0,28} & \multirow{2}{*}{0,07} & \multirow{2}{*}{\normalsize 57\%} \\
    \cline{2-8}       
    & \multirow{2}[2]{*}{KLIEP} & LOSO & \normalsize 33\% & \normalsize 0,64 & \normalsize 0,19 & \normalsize 0,16 & \normalsize 48\% \\
    &    & LOUO & \normalsize 32\% & \normalsize 0,68 & \normalsize 0,22 & \normalsize 0,06 & \normalsize 51\% \\
    \cline{2-8}      
    & \multirow{2}[2]{*}{TrAdaBoost} & LOSO & \normalsize 44\% & \normalsize 0,68 & \normalsize 0,25 & \normalsize 0,39 & \normalsize 56\% \\
    &    & LOUO & \normalsize 85\% & \normalsize 0,91 & \normalsize 0,90 & \normalsize 0,74 & \normalsize 88\% \\
    \bottomrule
    \bottomrule
    \end{tabular}%
  \renewcommand{\arraystretch}{1}
  \label{tab:performance_DA_ita}%
\end{table}%

The second part of our study focused on speech sentiment recognition when multi-language-corpus datasets are taken into account. The trials tested for this analysis are summarized in Table \ref{tab:expe_EMOVO}. Two different datasets were used: the English dataset CREMA-D-ADULT, used to train the model, and the Italian dataset EMOVO, as Target and Test set. Furthermore, similarly to elderly, the results obtained varying the domain adaptation technique (KLIEP and TrAdaBoost) and evaluation strategy (LOSO, LOUO) were compared with the performance reached by the classification model trained without domain adaptation. The values of accuracy, macro-F1 score and per-class F1-scores achieved in the different experiments are reported in Table \ref{tab:performance_DA_ita}. 
From the analysis of the results, it emerges how the best performances in both the validation strategies were obtained applying the TrAdaBoost domain adaptation method. In particular, the two macro F1-score values of 44\% and 85\% generated using respectively LOSO and LOUO validation strategies outperform the value of 35\% reached when no domain adaptation is considered. Similarly to elderly, the lowest general performances were instead reached applying the KLIEP domain adaptation strategy with macro F1-score values near to 32\% in both the analysis performed. 
Another general consideration regards the single classes recognition. In almost all the trials, the use of domain adaptation techniques allowed to better recognize the instances of Positive class, reaching often more balanced classification performances in identify the three sentiments. Nevertheless, the Negative sentiment is still the class better recognized from all the classification models examined, thus confirming what has already been observed on the elderly analysis.

Finally, the last remark concerns the performance differences between the two validation strategies applied. In particular, the partition of Target and Test set using utterances allows to achieve better results than the one based on subjects. This can be explained by the fact that, in addition to language, the division by utterance also takes into account the difference between people with regard to personal vocal characteristics or how they express their emotions.
Using this method, data from each of the analyzed subjects appear in each of the folds generated, allowing the classification model to better learn about vocal timbre differences or differences in the individuals' personalities.
However, it is worth to underline that both the datasets analyzed are acted, making perhaps more similar how the same subject expresses the same emotion, also in different sentences. For this reason, in future analyzes, it may be necessary to validate the hypotheses here proposed on new natural datasets collected in real situations.

\section{Conclusion}

The sentiment emotion recognition task is still an open field of research, especially when considering different languages and ages. 
In particular in the case of our interest, Italian elderly, no datasets are available in the literature. Domain adaptation techniques could partially solve this lack of data. However our preliminary results indicate that there is the urgency of a more realistic collection of data, that also faces the need of considering different ages. 
Domain adaptation techniques seem to better perform in case of cross-language datasets, paving the way for further researches in this direction. For what concerns the lack of performance increase applying domain adaptation models in the case of multi-age corpus, conclusions can not be drawn, due to the peculiarity of the datasets available (where the collected speeches were recorded by professional actors) and given the low presence of elderly people.   

\begin{acknowledgments}
  This research is supported by the FONDAZIONE CARIPLO “AMPEL: Artificial intelligence facing Multidimensional Poverty in ELderly” (Ref. 2020-0232). 
\end{acknowledgments}

%%
%% Define the bibliography file to be used
\bibliography{biblio}

\begin{thebibliography}{60}
\expandafter\ifx\csname natexlab\endcsname\relax\def\natexlab#1{#1}\fi
\providecommand{\url}[1]{\texttt{#1}}
\providecommand{\href}[2]{#2}
\providecommand{\path}[1]{#1}
\providecommand{\DOIprefix}{doi:}
\providecommand{\ArXivprefix}{arXiv:}
\providecommand{\URLprefix}{URL: }
\providecommand{\Pubmedprefix}{pmid:}
\providecommand{\doi}[1]{\href{http://dx.doi.org/#1}{\path{#1}}}
\providecommand{\Pubmed}[1]{\href{pmid:#1}{\path{#1}}}
\providecommand{\bibinfo}[2]{#2}
\ifx\xfnm\relax \def\xfnm[#1]{\unskip,\space#1}\fi
%Type = Article
\bibitem[{Lange et~al.(2022)Lange, Heerdink, and Van~Kleef}]{lange2022reading}
\bibinfo{author}{J.~Lange}, \bibinfo{author}{M.~W. Heerdink},
  \bibinfo{author}{G.~A. Van~Kleef},
\newblock \bibinfo{title}{Reading emotions, reading people: Emotion perception
  and inferences drawn from perceived emotions},
\newblock \bibinfo{journal}{Current Opinion in Psychology} \bibinfo{volume}{43}
  (\bibinfo{year}{2022}) \bibinfo{pages}{85--90}.
%Type = Article
\bibitem[{El~Ayadi et~al.(2011)El~Ayadi, Kamel, and Karray}]{el2011survey}
\bibinfo{author}{M.~El~Ayadi}, \bibinfo{author}{M.~S. Kamel},
  \bibinfo{author}{F.~Karray},
\newblock \bibinfo{title}{Survey on speech emotion recognition: Features,
  classification schemes, and databases},
\newblock \bibinfo{journal}{Pattern recognition} \bibinfo{volume}{44}
  (\bibinfo{year}{2011}) \bibinfo{pages}{572--587}.
%Type = Article
\bibitem[{Swain et~al.(2018)Swain, Routray, and
  Kabisatpathy}]{swain2018databases}
\bibinfo{author}{M.~Swain}, \bibinfo{author}{A.~Routray},
  \bibinfo{author}{P.~Kabisatpathy},
\newblock \bibinfo{title}{Databases, features and classifiers for speech
  emotion recognition: a review},
\newblock \bibinfo{journal}{International Journal of Speech Technology}
  \bibinfo{volume}{21} (\bibinfo{year}{2018}) \bibinfo{pages}{93--120}.
%Type = Article
\bibitem[{Anagnostopoulos et~al.(2015)Anagnostopoulos, Iliou, and
  Giannoukos}]{anagnostopoulos2015features}
\bibinfo{author}{C.-N. Anagnostopoulos}, \bibinfo{author}{T.~Iliou},
  \bibinfo{author}{I.~Giannoukos},
\newblock \bibinfo{title}{Features and classifiers for emotion recognition from
  speech: a survey from 2000 to 2011},
\newblock \bibinfo{journal}{Artificial Intelligence Review}
  \bibinfo{volume}{43} (\bibinfo{year}{2015}) \bibinfo{pages}{155--177}.
%Type = Article
\bibitem[{Wu and Zhang(2022)}]{wu2022design}
\bibinfo{author}{X.~Wu}, \bibinfo{author}{Q.~Zhang},
\newblock \bibinfo{title}{Design of aging smart home products based on radial
  basis function speech emotion recognition.},
\newblock \bibinfo{journal}{Frontiers in Psychology} \bibinfo{volume}{13}
  (\bibinfo{year}{2022}) \bibinfo{pages}{882709--882709}.
%Type = Article
\bibitem[{Ak{\c{c}}ay and O{\u{g}}uz(2020)}]{akccay2020speech}
\bibinfo{author}{M.~B. Ak{\c{c}}ay}, \bibinfo{author}{K.~O{\u{g}}uz},
\newblock \bibinfo{title}{Speech emotion recognition: Emotional models,
  databases, features, preprocessing methods, supporting modalities, and
  classifiers},
\newblock \bibinfo{journal}{Speech Communication} \bibinfo{volume}{116}
  (\bibinfo{year}{2020}) \bibinfo{pages}{56--76}.
%Type = Article
\bibitem[{France et~al.(2000)France, Shiavi, Silverman, Silverman, and
  Wilkes}]{france2000acoustical}
\bibinfo{author}{D.~J. France}, \bibinfo{author}{R.~G. Shiavi},
  \bibinfo{author}{S.~Silverman}, \bibinfo{author}{M.~Silverman},
  \bibinfo{author}{M.~Wilkes},
\newblock \bibinfo{title}{Acoustical properties of speech as indicators of
  depression and suicidal risk},
\newblock \bibinfo{journal}{IEEE transactions on Biomedical Engineering}
  \bibinfo{volume}{47} (\bibinfo{year}{2000}) \bibinfo{pages}{829--837}.
%Type = Inproceedings
\bibitem[{Hua et~al.(2006)Hua, Litman, Forbes-Riley, Rotaru, Tetreault, and
  Purandare}]{hua2006using}
\bibinfo{author}{A.~Hua}, \bibinfo{author}{D.~J. Litman},
  \bibinfo{author}{K.~Forbes-Riley}, \bibinfo{author}{M.~Rotaru},
  \bibinfo{author}{J.~Tetreault}, \bibinfo{author}{A.~Purandare},
\newblock \bibinfo{title}{Using system and user performance features to improve
  emotion detection in spoken tutoring dialogs},
\newblock in: \bibinfo{booktitle}{Proceedings of the Annual Conference of the
  International Speech Communication Association, INTERSPEECH},
  volume~\bibinfo{volume}{2}, \bibinfo{year}{2006}, pp.
  \bibinfo{pages}{797--800}.
%Type = Article
\bibitem[{Cevher et~al.(2019)Cevher, Zepf, and Klinger}]{cevher2019towards}
\bibinfo{author}{D.~Cevher}, \bibinfo{author}{S.~Zepf},
  \bibinfo{author}{R.~Klinger},
\newblock \bibinfo{title}{Towards multimodal emotion recognition in german
  speech events in cars using transfer learning},
\newblock \bibinfo{journal}{arXiv preprint arXiv:1909.02764}
  (\bibinfo{year}{2019}).
%Type = Inproceedings
\bibitem[{Nakatsu et~al.(1999)Nakatsu, Nicholson, and
  Tosa}]{nakatsu1999emotion}
\bibinfo{author}{R.~Nakatsu}, \bibinfo{author}{J.~Nicholson},
  \bibinfo{author}{N.~Tosa},
\newblock \bibinfo{title}{Emotion recognition and its application to computer
  agents with spontaneous interactive capabilities},
\newblock in: \bibinfo{booktitle}{Proceedings of the seventh ACM international
  conference on Multimedia (Part 1)}, \bibinfo{year}{1999}, pp.
  \bibinfo{pages}{343--351}.
%Type = Inproceedings
\bibitem[{Alhargan et~al.(2017)Alhargan, Cooke, and
  Binjammaz}]{alhargan2017multimodal}
\bibinfo{author}{A.~Alhargan}, \bibinfo{author}{N.~Cooke},
  \bibinfo{author}{T.~Binjammaz},
\newblock \bibinfo{title}{Multimodal affect recognition in an interactive
  gaming environment using eye tracking and speech signals},
\newblock in: \bibinfo{booktitle}{Proceedings of the 19th ACM international
  conference on multimodal interaction}, \bibinfo{year}{2017}, pp.
  \bibinfo{pages}{479--486}.
%Type = Article
\bibitem[{Low et~al.(2010)Low, Maddage, Lech, Sheeber, and
  Allen}]{low2010detection}
\bibinfo{author}{L.-S.~A. Low}, \bibinfo{author}{N.~C. Maddage},
  \bibinfo{author}{M.~Lech}, \bibinfo{author}{L.~B. Sheeber},
  \bibinfo{author}{N.~B. Allen},
\newblock \bibinfo{title}{Detection of clinical depression in adolescents’
  speech during family interactions},
\newblock \bibinfo{journal}{IEEE Transactions on Biomedical Engineering}
  \bibinfo{volume}{58} (\bibinfo{year}{2010}) \bibinfo{pages}{574--586}.
%Type = Inproceedings
\bibitem[{Al~Machot et~al.(2011)Al~Machot, Mosa, Dabbour, Fasih,
  Schwarzlm{\"u}ller, Ali, and Kyamakya}]{al2011novel}
\bibinfo{author}{F.~Al~Machot}, \bibinfo{author}{A.~H. Mosa},
  \bibinfo{author}{K.~Dabbour}, \bibinfo{author}{A.~Fasih},
  \bibinfo{author}{C.~Schwarzlm{\"u}ller}, \bibinfo{author}{M.~Ali},
  \bibinfo{author}{K.~Kyamakya},
\newblock \bibinfo{title}{A novel real-time emotion detection system from audio
  streams based on bayesian quadratic discriminate classifier for adas},
\newblock in: \bibinfo{booktitle}{Proceedings of the Joint INDS'11 \&
  ISTET'11}, \bibinfo{organization}{IEEE}, \bibinfo{year}{2011}, pp.
  \bibinfo{pages}{1--5}.
%Type = Inproceedings
\bibitem[{Gupta and Rajput(2007)}]{gupta2007two}
\bibinfo{author}{P.~Gupta}, \bibinfo{author}{N.~Rajput},
\newblock \bibinfo{title}{Two-stream emotion recognition for call center
  monitoring},
\newblock in: \bibinfo{booktitle}{Eighth Annual Conference of the International
  Speech Communication Association}, \bibinfo{organization}{Citeseer},
  \bibinfo{year}{2007}.
%Type = Inproceedings
\bibitem[{Vaudable and Devillers(2012)}]{vaudable2012negative}
\bibinfo{author}{C.~Vaudable}, \bibinfo{author}{L.~Devillers},
\newblock \bibinfo{title}{Negative emotions detection as an indicator of
  dialogs quality in call centers},
\newblock in: \bibinfo{booktitle}{2012 IEEE International Conference on
  Acoustics, Speech and Signal Processing (ICASSP)},
  \bibinfo{organization}{IEEE}, \bibinfo{year}{2012}, pp.
  \bibinfo{pages}{5109--5112}.
%Type = Inproceedings
\bibitem[{Hegel et~al.(2006)Hegel, Spexard, Wrede, Horstmann, and
  Vogt}]{hegel2006playing}
\bibinfo{author}{F.~Hegel}, \bibinfo{author}{T.~Spexard},
  \bibinfo{author}{B.~Wrede}, \bibinfo{author}{G.~Horstmann},
  \bibinfo{author}{T.~Vogt},
\newblock \bibinfo{title}{Playing a different imitation game: Interaction with
  an empathic android robot},
\newblock in: \bibinfo{booktitle}{2006 6th IEEE-RAS International Conference on
  Humanoid Robots}, \bibinfo{organization}{IEEE}, \bibinfo{year}{2006}, pp.
  \bibinfo{pages}{56--61}.
%Type = Incollection
\bibitem[{Jones and Deeming(2008)}]{jones2008affective}
\bibinfo{author}{C.~Jones}, \bibinfo{author}{A.~Deeming},
\newblock \bibinfo{title}{Affective human-robotic interaction},
\newblock in: \bibinfo{booktitle}{Affect and emotion in human-computer
  interaction}, \bibinfo{publisher}{Springer}, \bibinfo{year}{2008}, pp.
  \bibinfo{pages}{175--185}.
%Type = Article
\bibitem[{Fahad et~al.(2021)Fahad, Ranjan, Yadav, and Deepak}]{fahad2021survey}
\bibinfo{author}{M.~S. Fahad}, \bibinfo{author}{A.~Ranjan},
  \bibinfo{author}{J.~Yadav}, \bibinfo{author}{A.~Deepak},
\newblock \bibinfo{title}{A survey of speech emotion recognition in natural
  environment},
\newblock \bibinfo{journal}{Digital Signal Processing} \bibinfo{volume}{110}
  (\bibinfo{year}{2021}) \bibinfo{pages}{102951}.
%Type = Article
\bibitem[{Atmaja et~al.(2022)Atmaja, Sasou, and Akagi}]{atmaja2022survey}
\bibinfo{author}{B.~T. Atmaja}, \bibinfo{author}{A.~Sasou},
  \bibinfo{author}{M.~Akagi},
\newblock \bibinfo{title}{Survey on bimodal speech emotion recognition from
  acoustic and linguistic information fusion},
\newblock \bibinfo{journal}{Speech Communication}  (\bibinfo{year}{2022}).
%Type = Article
\bibitem[{Thakur and Dhull(2021)}]{thakur2021speech}
\bibinfo{author}{A.~Thakur}, \bibinfo{author}{S.~Dhull},
\newblock \bibinfo{title}{Speech emotion recognition: A review},
\newblock \bibinfo{journal}{Advances in Communication and Computational
  Technology}  (\bibinfo{year}{2021}) \bibinfo{pages}{815--827}.
%Type = Inproceedings
\bibitem[{Stolar et~al.(2017)Stolar, Lech, Bolia, and Skinner}]{stolar2017real}
\bibinfo{author}{M.~N. Stolar}, \bibinfo{author}{M.~Lech},
  \bibinfo{author}{R.~S. Bolia}, \bibinfo{author}{M.~Skinner},
\newblock \bibinfo{title}{Real time speech emotion recognition using rgb image
  classification and transfer learning},
\newblock in: \bibinfo{booktitle}{2017 11th International Conference on Signal
  Processing and Communication Systems (ICSPCS)}, \bibinfo{organization}{IEEE},
  \bibinfo{year}{2017}, pp. \bibinfo{pages}{1--8}.
%Type = Inproceedings
\bibitem[{Boateng and Kowatsch(2020)}]{boateng2020speech}
\bibinfo{author}{G.~Boateng}, \bibinfo{author}{T.~Kowatsch},
\newblock \bibinfo{title}{Speech emotion recognition among elderly individuals
  using multimodal fusion and transfer learning},
\newblock in: \bibinfo{booktitle}{Companion Publication of the 2020
  International Conference on Multimodal Interaction}, \bibinfo{year}{2020},
  pp. \bibinfo{pages}{12--16}.
%Type = Article
\bibitem[{Swain et~al.(2015)Swain, Sahoo, Routray, Kabisatpathy, and
  Kundu}]{swain2015study}
\bibinfo{author}{M.~Swain}, \bibinfo{author}{S.~Sahoo},
  \bibinfo{author}{A.~Routray}, \bibinfo{author}{P.~Kabisatpathy},
  \bibinfo{author}{J.~N. Kundu},
\newblock \bibinfo{title}{Study of feature combination using hmm and svm for
  multilingual odiya speech emotion recognition},
\newblock \bibinfo{journal}{International Journal of Speech Technology}
  \bibinfo{volume}{18} (\bibinfo{year}{2015}) \bibinfo{pages}{387--393}.
%Type = Inproceedings
\bibitem[{Latif et~al.(2018)Latif, Qayyum, Usman, and Qadir}]{latif2018cross}
\bibinfo{author}{S.~Latif}, \bibinfo{author}{A.~Qayyum},
  \bibinfo{author}{M.~Usman}, \bibinfo{author}{J.~Qadir},
\newblock \bibinfo{title}{Cross lingual speech emotion recognition: Urdu vs.
  western languages},
\newblock in: \bibinfo{booktitle}{2018 International Conference on Frontiers of
  Information Technology (FIT)}, \bibinfo{organization}{IEEE},
  \bibinfo{year}{2018}, pp. \bibinfo{pages}{88--93}.
%Type = Article
\bibitem[{Wani et~al.(2021)Wani, Gunawan, Qadri, Kartiwi, and
  Ambikairajah}]{wani2021comprehensive}
\bibinfo{author}{T.~M. Wani}, \bibinfo{author}{T.~S. Gunawan},
  \bibinfo{author}{S.~A.~A. Qadri}, \bibinfo{author}{M.~Kartiwi},
  \bibinfo{author}{E.~Ambikairajah},
\newblock \bibinfo{title}{A comprehensive review of speech emotion recognition
  systems},
\newblock \bibinfo{journal}{IEEE Access} \bibinfo{volume}{9}
  (\bibinfo{year}{2021}) \bibinfo{pages}{47795--47814}.
%Type = Inproceedings
\bibitem[{Lugger et~al.(2009)Lugger, Janoir, and Yang}]{lugger2009combining}
\bibinfo{author}{M.~Lugger}, \bibinfo{author}{M.-E. Janoir},
  \bibinfo{author}{B.~Yang},
\newblock \bibinfo{title}{Combining classifiers with diverse feature sets for
  robust speaker independent emotion recognition},
\newblock in: \bibinfo{booktitle}{2009 17th European Signal Processing
  Conference}, \bibinfo{organization}{IEEE}, \bibinfo{year}{2009}, pp.
  \bibinfo{pages}{1225--1229}.
%Type = Inproceedings
\bibitem[{Schuller et~al.(2005)Schuller, Lang, and Rigoll}]{schuller2005robust}
\bibinfo{author}{B.~Schuller}, \bibinfo{author}{M.~Lang},
  \bibinfo{author}{G.~Rigoll},
\newblock \bibinfo{title}{Robust acoustic speech emotion recognition by
  ensembles of classifiers},
\newblock in: \bibinfo{booktitle}{Tagungsband Fortschritte der Akustik-DAGA\#
  05, M{\"u}nchen}, \bibinfo{year}{2005}.
%Type = Inproceedings
\bibitem[{Badshah et~al.(2017)Badshah, Ahmad, Rahim, and
  Baik}]{badshah2017speech}
\bibinfo{author}{A.~M. Badshah}, \bibinfo{author}{J.~Ahmad},
  \bibinfo{author}{N.~Rahim}, \bibinfo{author}{S.~W. Baik},
\newblock \bibinfo{title}{Speech emotion recognition from spectrograms with
  deep convolutional neural network},
\newblock in: \bibinfo{booktitle}{2017 international conference on platform
  technology and service (PlatCon)}, \bibinfo{organization}{IEEE},
  \bibinfo{year}{2017}, pp. \bibinfo{pages}{1--5}.
%Type = Article
\bibitem[{Aghajani and Esmaili Paeen~Afrakoti(2020)}]{aghajani2020speech}
\bibinfo{author}{K.~Aghajani}, \bibinfo{author}{I.~Esmaili Paeen~Afrakoti},
\newblock \bibinfo{title}{Speech emotion recognition using scalogram based deep
  structure},
\newblock \bibinfo{journal}{International Journal of Engineering}
  \bibinfo{volume}{33} (\bibinfo{year}{2020}) \bibinfo{pages}{285--292}.
%Type = Article
\bibitem[{Cho et~al.(2019)Cho, Pappagari, Kulkarni, Villalba, Carmiel, and
  Dehak}]{cho2019deep}
\bibinfo{author}{J.~Cho}, \bibinfo{author}{R.~Pappagari},
  \bibinfo{author}{P.~Kulkarni}, \bibinfo{author}{J.~Villalba},
  \bibinfo{author}{Y.~Carmiel}, \bibinfo{author}{N.~Dehak},
\newblock \bibinfo{title}{Deep neural networks for emotion recognition
  combining audio and transcripts},
\newblock \bibinfo{journal}{arXiv preprint arXiv:1911.00432}
  (\bibinfo{year}{2019}).
%Type = Inproceedings
\bibitem[{Kumbhar and Bhandari(2019)}]{kumbhar2019speech}
\bibinfo{author}{H.~S. Kumbhar}, \bibinfo{author}{S.~U. Bhandari},
\newblock \bibinfo{title}{Speech emotion recognition using mfcc features and
  lstm network},
\newblock in: \bibinfo{booktitle}{2019 5th International Conference On
  Computing, Communication, Control And Automation (ICCUBEA)},
  \bibinfo{organization}{IEEE}, \bibinfo{year}{2019}, pp.
  \bibinfo{pages}{1--3}.
%Type = Inproceedings
\bibitem[{Atmaja and Akagi(2019)}]{atmaja2019speech}
\bibinfo{author}{B.~T. Atmaja}, \bibinfo{author}{M.~Akagi},
\newblock \bibinfo{title}{Speech emotion recognition based on speech segment
  using lstm with attention model},
\newblock in: \bibinfo{booktitle}{2019 IEEE International Conference on Signals
  and Systems (ICSigSys)}, \bibinfo{organization}{IEEE}, \bibinfo{year}{2019},
  pp. \bibinfo{pages}{40--44}.
%Type = Inproceedings
\bibitem[{Jian et~al.(2022)Jian, Xiang, and Huang}]{jian2022speech}
\bibinfo{author}{Q.~Jian}, \bibinfo{author}{M.~Xiang},
  \bibinfo{author}{W.~Huang},
\newblock \bibinfo{title}{A speech emotion recognition method for the elderly
  based on feature fusion and attention mechanism},
\newblock in: \bibinfo{booktitle}{Third International Conference on Electronics
  and Communication; Network and Computer Technology (ECNCT 2021)}, volume
  \bibinfo{volume}{12167}, \bibinfo{organization}{SPIE}, \bibinfo{year}{2022},
  pp. \bibinfo{pages}{398--403}.
%Type = Inproceedings
\bibitem[{Neumann and Vu(2019)}]{neumann2019improving}
\bibinfo{author}{M.~Neumann}, \bibinfo{author}{N.~T. Vu},
\newblock \bibinfo{title}{Improving speech emotion recognition with
  unsupervised representation learning on unlabeled speech},
\newblock in: \bibinfo{booktitle}{ICASSP 2019-2019 IEEE International
  Conference on Acoustics, Speech and Signal Processing (ICASSP)},
  \bibinfo{organization}{IEEE}, \bibinfo{year}{2019}, pp.
  \bibinfo{pages}{7390--7394}.
%Type = Article
\bibitem[{Koolagudi and Rao(2012)}]{koolagudi2012emotion}
\bibinfo{author}{S.~G. Koolagudi}, \bibinfo{author}{K.~S. Rao},
\newblock \bibinfo{title}{Emotion recognition from speech: a review},
\newblock \bibinfo{journal}{International journal of speech technology}
  \bibinfo{volume}{15} (\bibinfo{year}{2012}) \bibinfo{pages}{99--117}.
%Type = Inproceedings
\bibitem[{Ringeval et~al.(2013)Ringeval, Sonderegger, Sauer, and
  Lalanne}]{ringeval2013introducing}
\bibinfo{author}{F.~Ringeval}, \bibinfo{author}{A.~Sonderegger},
  \bibinfo{author}{J.~Sauer}, \bibinfo{author}{D.~Lalanne},
\newblock \bibinfo{title}{Introducing the recola multimodal corpus of remote
  collaborative and affective interactions},
\newblock in: \bibinfo{booktitle}{2013 10th IEEE international conference and
  workshops on automatic face and gesture recognition (FG)},
  \bibinfo{organization}{IEEE}, \bibinfo{year}{2013}, pp.
  \bibinfo{pages}{1--8}.
%Type = Book
\bibitem[{Steidl(2009)}]{steidl2009automatic}
\bibinfo{author}{S.~Steidl}, \bibinfo{title}{Automatic classification of
  emotion related user states in spontaneous children's speech},
  \bibinfo{publisher}{Logos-Verlag Berlin, Germany}, \bibinfo{year}{2009}.
%Type = Inproceedings
\bibitem[{Fan et~al.(2021)Fan, Xu, Xing, Chen, and Huang}]{fan2021lssed}
\bibinfo{author}{W.~Fan}, \bibinfo{author}{X.~Xu}, \bibinfo{author}{X.~Xing},
  \bibinfo{author}{W.~Chen}, \bibinfo{author}{D.~Huang},
\newblock \bibinfo{title}{Lssed: a large-scale dataset and benchmark for speech
  emotion recognition},
\newblock in: \bibinfo{booktitle}{ICASSP 2021-2021 IEEE International
  Conference on Acoustics, Speech and Signal Processing (ICASSP)},
  \bibinfo{organization}{IEEE}, \bibinfo{year}{2021}, pp.
  \bibinfo{pages}{641--645}.
%Type = Article
\bibitem[{Morrison et~al.(2007)Morrison, Wang, and
  De~Silva}]{morrison2007ensemble}
\bibinfo{author}{D.~Morrison}, \bibinfo{author}{R.~Wang},
  \bibinfo{author}{L.~C. De~Silva},
\newblock \bibinfo{title}{Ensemble methods for spoken emotion recognition in
  call-centres},
\newblock \bibinfo{journal}{Speech communication} \bibinfo{volume}{49}
  (\bibinfo{year}{2007}) \bibinfo{pages}{98--112}.
%Type = Article
\bibitem[{Parada-Cabaleiro et~al.(2018)Parada-Cabaleiro, Costantini, Batliner,
  Baird, and Schuller}]{parada2018categorical}
\bibinfo{author}{E.~Parada-Cabaleiro}, \bibinfo{author}{G.~Costantini},
  \bibinfo{author}{A.~Batliner}, \bibinfo{author}{A.~Baird},
  \bibinfo{author}{B.~Schuller},
\newblock \bibinfo{title}{Categorical vs dimensional perception of italian
  emotional speech}  (\bibinfo{year}{2018}).
%Type = Inproceedings
\bibitem[{Hozjan et~al.(2002)Hozjan, Kacic, Moreno, Bonafonte, and
  Nogueiras}]{hozjan2002interface}
\bibinfo{author}{V.~Hozjan}, \bibinfo{author}{Z.~Kacic},
  \bibinfo{author}{A.~Moreno}, \bibinfo{author}{A.~Bonafonte},
  \bibinfo{author}{A.~Nogueiras},
\newblock \bibinfo{title}{Interface databases: Design and collection of a
  multilingual emotional speech database.},
\newblock in: \bibinfo{booktitle}{LREC}, \bibinfo{year}{2002}.
%Type = Article
\bibitem[{Deliyski(2001)}]{deliyski2001effects}
\bibinfo{author}{D.~Deliyski, Steve An~Xue},
\newblock \bibinfo{title}{Effects of aging on selected acoustic voice
  parameters: Preliminary normative data and educational implications},
\newblock \bibinfo{journal}{Educational gerontology} \bibinfo{volume}{27}
  (\bibinfo{year}{2001}) \bibinfo{pages}{159--168}.
%Type = Article
\bibitem[{Sundberg et~al.(1998)Sundberg, Th{\"o}rnvik, and
  S{\"o}derstr{\"o}m}]{sundberg1998age}
\bibinfo{author}{J.~Sundberg}, \bibinfo{author}{M.~N. Th{\"o}rnvik},
  \bibinfo{author}{A.~M. S{\"o}derstr{\"o}m},
\newblock \bibinfo{title}{Age and voice quality in professional singers},
\newblock \bibinfo{journal}{Logopedics Phoniatrics Vocology}
  \bibinfo{volume}{23} (\bibinfo{year}{1998}) \bibinfo{pages}{169--176}.
%Type = Inproceedings
\bibitem[{Verma and Mukhopadhyay(2016)}]{verma2016age}
\bibinfo{author}{D.~Verma}, \bibinfo{author}{D.~Mukhopadhyay},
\newblock \bibinfo{title}{Age driven automatic speech emotion recognition
  system},
\newblock in: \bibinfo{booktitle}{2016 International Conference on Computing,
  Communication and Automation (ICCCA)}, \bibinfo{organization}{IEEE},
  \bibinfo{year}{2016}, pp. \bibinfo{pages}{1005--1010}.
%Type = Article
\bibitem[{So{\u{g}}anc{\i}o{\u{g}}lu et~al.(2020)So{\u{g}}anc{\i}o{\u{g}}lu,
  Verkholyak, Kaya, Fedotov, Cad{\'e}e, Salah, and
  Karpov}]{souganciouglu2020everything}
\bibinfo{author}{G.~So{\u{g}}anc{\i}o{\u{g}}lu},
  \bibinfo{author}{O.~Verkholyak}, \bibinfo{author}{H.~Kaya},
  \bibinfo{author}{D.~Fedotov}, \bibinfo{author}{T.~Cad{\'e}e},
  \bibinfo{author}{A.~A. Salah}, \bibinfo{author}{A.~Karpov},
\newblock \bibinfo{title}{Is everything fine, grandma? acoustic and linguistic
  modeling for robust elderly speech emotion recognition},
\newblock \bibinfo{journal}{arXiv preprint arXiv:2009.03432}
  (\bibinfo{year}{2020}).
%Type = Article
\bibitem[{Schuller et~al.(2020)Schuller, Batliner, Bergler, Messner, Hamilton,
  Amiriparian, Baird, Rizos, Schmitt, Stappen et~al.}]{schuller2020interspeech}
\bibinfo{author}{B.~W. Schuller}, \bibinfo{author}{A.~Batliner},
  \bibinfo{author}{C.~Bergler}, \bibinfo{author}{E.-M. Messner},
  \bibinfo{author}{A.~Hamilton}, \bibinfo{author}{S.~Amiriparian},
  \bibinfo{author}{A.~Baird}, \bibinfo{author}{G.~Rizos},
  \bibinfo{author}{M.~Schmitt}, \bibinfo{author}{L.~Stappen}, et~al.,
\newblock \bibinfo{title}{The interspeech 2020 computational paralinguistics
  challenge: Elderly emotion, breathing \& masks}  (\bibinfo{year}{2020}).
%Type = Article
\bibitem[{Cao et~al.(2014)Cao, Cooper, Keutmann, Gur, Nenkova, and
  Verma}]{cao2014crema}
\bibinfo{author}{H.~Cao}, \bibinfo{author}{D.~G. Cooper},
  \bibinfo{author}{M.~K. Keutmann}, \bibinfo{author}{R.~C. Gur},
  \bibinfo{author}{A.~Nenkova}, \bibinfo{author}{R.~Verma},
\newblock \bibinfo{title}{Crema-d: Crowd-sourced emotional multimodal actors
  dataset},
\newblock \bibinfo{journal}{IEEE transactions on affective computing}
  \bibinfo{volume}{5} (\bibinfo{year}{2014}) \bibinfo{pages}{377--390}.
%Type = Misc
\bibitem[{Pichora-Fuller and Dupuis(2020)}]{SP2/E8H2MF_2020}
\bibinfo{author}{M.~K. Pichora-Fuller}, \bibinfo{author}{K.~Dupuis},
  \bibinfo{title}{{Toronto emotional speech set (TESS)}}, \bibinfo{year}{2020}.
  \URLprefix \url{https://doi.org/10.5683/SP2/E8H2MF}.
  \DOIprefix\doi{10.5683/SP2/E8H2MF}.
%Type = Inproceedings
\bibitem[{Costantini et~al.(2014)Costantini, Iaderola, Paoloni, and
  Todisco}]{costantini2014emovo}
\bibinfo{author}{G.~Costantini}, \bibinfo{author}{I.~Iaderola},
  \bibinfo{author}{A.~Paoloni}, \bibinfo{author}{M.~Todisco},
\newblock \bibinfo{title}{Emovo corpus: an italian emotional speech database},
\newblock in: \bibinfo{booktitle}{International Conference on Language
  Resources and Evaluation (LREC 2014)}, \bibinfo{organization}{European
  Language Resources Association (ELRA)}, \bibinfo{year}{2014}, pp.
  \bibinfo{pages}{3501--3504}.
%Type = Article
\bibitem[{Poria et~al.(2018)Poria, Hazarika, Majumder, Naik, Cambria, and
  Mihalcea}]{poria2018meld}
\bibinfo{author}{S.~Poria}, \bibinfo{author}{D.~Hazarika},
  \bibinfo{author}{N.~Majumder}, \bibinfo{author}{G.~Naik},
  \bibinfo{author}{E.~Cambria}, \bibinfo{author}{R.~Mihalcea},
\newblock \bibinfo{title}{Meld: A multimodal multi-party dataset for emotion
  recognition in conversations},
\newblock \bibinfo{journal}{arXiv preprint arXiv:1810.02508}
  (\bibinfo{year}{2018}).
%Type = Inproceedings
\bibitem[{Zadeh et~al.(2018)Zadeh, Liang, Poria, Cambria, and
  Morency}]{zadeh2018multimodal}
\bibinfo{author}{A.~B. Zadeh}, \bibinfo{author}{P.~P. Liang},
  \bibinfo{author}{S.~Poria}, \bibinfo{author}{E.~Cambria},
  \bibinfo{author}{L.-P. Morency},
\newblock \bibinfo{title}{Multimodal language analysis in the wild: Cmu-mosei
  dataset and interpretable dynamic fusion graph},
\newblock in: \bibinfo{booktitle}{Proceedings of the 56th Annual Meeting of the
  Association for Computational Linguistics (Volume 1: Long Papers)},
  \bibinfo{year}{2018}, pp. \bibinfo{pages}{2236--2246}.
%Type = Inproceedings
\bibitem[{Chen and Guestrin(2016)}]{chen2016xgboost}
\bibinfo{author}{T.~Chen}, \bibinfo{author}{C.~Guestrin},
\newblock \bibinfo{title}{Xgboost: A scalable tree boosting system},
\newblock in: \bibinfo{booktitle}{Proceedings of the 22nd acm sigkdd
  international conference on knowledge discovery and data mining},
  \bibinfo{year}{2016}, pp. \bibinfo{pages}{785--794}.
%Type = Article
\bibitem[{Sugiyama et~al.(2007)Sugiyama, Nakajima, Kashima, Buenau, and
  Kawanabe}]{sugiyama2007direct}
\bibinfo{author}{M.~Sugiyama}, \bibinfo{author}{S.~Nakajima},
  \bibinfo{author}{H.~Kashima}, \bibinfo{author}{P.~Buenau},
  \bibinfo{author}{M.~Kawanabe},
\newblock \bibinfo{title}{Direct importance estimation with model selection and
  its application to covariate shift adaptation},
\newblock \bibinfo{journal}{Advances in neural information processing systems}
  \bibinfo{volume}{20} (\bibinfo{year}{2007}).
%Type = Inproceedings
\bibitem[{Dai et~al.(2007)Dai, Yang, Xue, and Yu}]{inproceedings}
\bibinfo{author}{W.~Dai}, \bibinfo{author}{Q.~Yang}, \bibinfo{author}{G.-R.
  Xue}, \bibinfo{author}{Y.~Yu},
\newblock \bibinfo{title}{Boosting for transfer learning},
\newblock volume \bibinfo{volume}{227}, \bibinfo{year}{2007}, pp.
  \bibinfo{pages}{193--200}. \DOIprefix\doi{10.1145/1273496.1273521}.
%Type = Article
\bibitem[{Grandini et~al.(2020)Grandini, Bagli, and
  Visani}]{grandini2020metrics}
\bibinfo{author}{M.~Grandini}, \bibinfo{author}{E.~Bagli},
  \bibinfo{author}{G.~Visani},
\newblock \bibinfo{title}{Metrics for multi-class classification: an overview},
\newblock \bibinfo{journal}{arXiv preprint arXiv:2008.05756}
  (\bibinfo{year}{2020}).
%Type = Inproceedings
\bibitem[{Lipton et~al.(2014)Lipton, Elkan, and
  Naryanaswamy}]{lipton2014optimal}
\bibinfo{author}{Z.~C. Lipton}, \bibinfo{author}{C.~Elkan},
  \bibinfo{author}{B.~Naryanaswamy},
\newblock \bibinfo{title}{Optimal thresholding of classifiers to maximize f1
  measure},
\newblock in: \bibinfo{booktitle}{Joint European Conference on Machine Learning
  and Knowledge Discovery in Databases}, \bibinfo{organization}{Springer},
  \bibinfo{year}{2014}, pp. \bibinfo{pages}{225--239}.
%Type = Article
\bibitem[{Birch et~al.(2021)Birch, Griffiths, and
  Morgan}]{birch2021environmental}
\bibinfo{author}{B.~Birch}, \bibinfo{author}{C.~Griffiths},
  \bibinfo{author}{A.~Morgan},
\newblock \bibinfo{title}{Environmental effects on reliability and accuracy of
  mfcc based voice recognition for industrial human-robot-interaction},
\newblock \bibinfo{journal}{Proceedings of the Institution of Mechanical
  Engineers, Part B: Journal of Engineering Manufacture} \bibinfo{volume}{235}
  (\bibinfo{year}{2021}) \bibinfo{pages}{1939--1948}.
%Type = Inproceedings
\bibitem[{Eyben et~al.(2010)Eyben, W{\"o}llmer, and
  Schuller}]{eyben2010opensmile}
\bibinfo{author}{F.~Eyben}, \bibinfo{author}{M.~W{\"o}llmer},
  \bibinfo{author}{B.~Schuller},
\newblock \bibinfo{title}{Opensmile: the munich versatile and fast open-source
  audio feature extractor},
\newblock in: \bibinfo{booktitle}{Proceedings of the 18th ACM international
  conference on Multimedia}, \bibinfo{year}{2010}, pp.
  \bibinfo{pages}{1459--1462}.
%Type = Article
\bibitem[{Eyben et~al.(2015)Eyben, Scherer, Schuller, Sundberg, Andr{\'e},
  Busso, Devillers, Epps, Laukka, Narayanan et~al.}]{eyben2015geneva}
\bibinfo{author}{F.~Eyben}, \bibinfo{author}{K.~R. Scherer},
  \bibinfo{author}{B.~W. Schuller}, \bibinfo{author}{J.~Sundberg},
  \bibinfo{author}{E.~Andr{\'e}}, \bibinfo{author}{C.~Busso},
  \bibinfo{author}{L.~Y. Devillers}, \bibinfo{author}{J.~Epps},
  \bibinfo{author}{P.~Laukka}, \bibinfo{author}{S.~S. Narayanan}, et~al.,
\newblock \bibinfo{title}{The geneva minimalistic acoustic parameter set
  (gemaps) for voice research and affective computing},
\newblock \bibinfo{journal}{IEEE transactions on affective computing}
  \bibinfo{volume}{7} (\bibinfo{year}{2015}) \bibinfo{pages}{190--202}.
%Type = Article
\bibitem[{Chawla et~al.(2002)Chawla, Bowyer, Hall, and
  Kegelmeyer}]{chawla2002smote}
\bibinfo{author}{N.~V. Chawla}, \bibinfo{author}{K.~W. Bowyer},
  \bibinfo{author}{L.~O. Hall}, \bibinfo{author}{W.~P. Kegelmeyer},
\newblock \bibinfo{title}{Smote: synthetic minority over-sampling technique},
\newblock \bibinfo{journal}{Journal of artificial intelligence research}
  \bibinfo{volume}{16} (\bibinfo{year}{2002}) \bibinfo{pages}{321--357}.

\end{thebibliography}

\appendix

\end{document}